\documentclass[runningheads]{llncs}
\usepackage[T1]{fontenc}
\usepackage{graphicx}
\usepackage{xspace}
\usepackage{cleveref}
\newcommand{\tool}{\textit{Evolaris}\xspace}

\begin{document}
\title{\tool: A Roadmap to Self-Evolving Software Intelligence Management}
%
%
\author{Chengwei Liu\inst{1} \and
Wenbo Guo\inst{1} \and
Yuxin Zhang\inst{2} \and
Limin Wang\inst{3} \and
Sen Chen\inst{2} \and
Lei Bu\inst{3} \and
Yang Liu\inst{1}
}
\authorrunning{C. Liu et al.}
%
\institute{Nanyang Technological University, Singapore 
\and
Nankai University, China \and
Nanjing University, China
}
\maketitle              

\begin{abstract}
In recent years, the landscape of software threats has become significantly more dynamic and distributed. Security vulnerabilities are no longer discovered and shared only through formal channels such as public vulnerability databases or vendor advisories. Increasingly, critical threat information emerges informally through blogs, social media, developer forums, open source repositories, and even underground communities. 
To this end, capturing such intelligence in a timely manner is essential for maintaining situational awareness and enabling prompt security responses. However, this remains a complex challenge due to the fragmented nature of data sources and the technical difficulty of collecting, parsing, mapping, and validating information at scale. 
To address this, we propose \tool, a self-evolving software intelligence system built on a multi-agent framework. \tool is designed to support a full-stack workflow, where agents operate independently but coordinate through shared context to perform tasks such as information discovery, reasoning, gap completion, validation, and risk detection. This architecture enables the platform to learn from new inputs, refine its internal knowledge, and adapt to emerging threat patterns over time, which could continuously improve the precision, timeliness, and scalability of software threat analysis, and offers a sustainable foundation for proactive security decision-making and strengthens the broader ecosystem of security threat understanding.

\end{abstract}

\section{Introduction}

The volume of software security threats has increased at an accelerating rate, placing substantial strain on the broader security infrastructure. Centralized validation entities, such as the National Vulnerability Database (NVD)~\cite{nvdweb} and vendor-coordinated disclosure programs, face mounting challenges in processing and curating threat reports with both timeliness and accuracy. This growing workload has contributed to observable deficiencies in data quality, including delayed publication, incomplete records, and inconsistent threat characterizations.
Concurrently, an increasing proportion of threat-related intelligence is being disseminated through unstructured and rapidly updated online platforms, including technical blogs, developer forums, community-maintained issue trackers, and social media. Although such sources are heterogeneous and less formally governed, they frequently contain early-stage indicators, technical analyses, and mitigation strategies that are absent from official repositories. 

To this end, existing researchers have investigated how to collect, interpret, and utilize these online resources in various ways. 
For instance, researchers have developed automated systems to gather threat intelligence from various sources, including the internet~\cite{mulwad2011extracting}, social media~\cite{sabottke2015vulnerability}, developer communities~\cite{purba2023extracting}, security forums, and code repositories\cite{bouwman2022helping}, some researchers also focus on natural language processing and knowledge graphs to automate cybersecurity knowledge extraction from unstructured text~\cite{jo2022vulcan}\cite{satvat2021extractor}\cite{gao2021system}.
However, there are still challenges yet to be solved, which hinder the real-world applicability of proposed solutions. 
First, the relevant information is scattered across heterogeneous sources with varying formats, update frequencies, and levels of trust, which still heavily rely on specialised designs that require human expertise. Second, the content is often embedded in unstructured natural language, lacking explicit structure or machine-readable identifiers to indicate semantics in the targeted context. Third, aligning such information with existing software projects, components, or ecosystem-level behaviors requires non-trivial contextual reasoning. Finally, assessing the reliability and significance of threat information remains difficult without deeper analysis and cross-validation.
These challenges are further intensified by the rapid emergence of new threat categories and shifts in how security intelligence is expressed and shared. As a result, there is a critical need for intelligent systems that can continuously monitor, extract, interpret, and reason over dynamic and diverse security information sources.

To address this gap, we propose \tool{}, a self-evolving software intelligence platform that automates the full lifecycle of threat information processing. \tool{} employs a collaborative multi-agent architecture in which specialized agents carry out key tasks, including data discovery, analytical reasoning, information completion, validation, and risk detection. These agents share context, refine intermediate outputs, and collectively evolve the system's internal models as new data becomes available.
\tool{} is designed to adapt to emerging forms of intelligence, support scalable analysis across multiple data sources, and improve its analytical capabilities over time. By automating reasoning and knowledge refinement in a distributed and evolving manner, \tool{} enhances the accuracy, timeliness, and usefulness of software security threat intelligence. Our goal is to establish a foundational platform that enables continuous, context-aware, and proactive software security understanding.

\section{Overview of \tool}

\begin{figure}[t]
    \centering
    \includegraphics[width=0.95\textwidth]{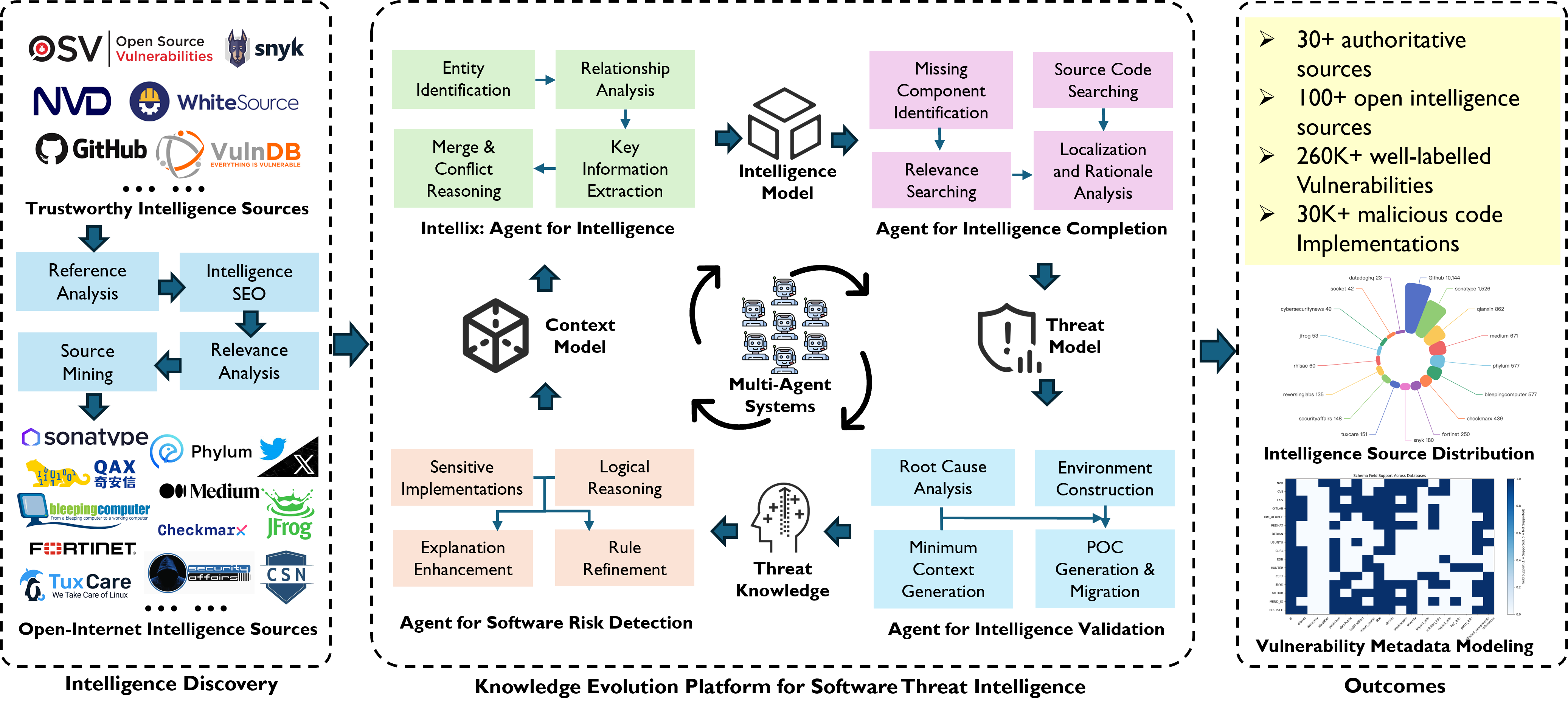}
    \caption{Overviews of \tool}
    \label{fig:overviews}
\end{figure}

\Cref{fig:overviews} presents the overview roadmap of \tool. Specifically, it contains two major parts:
    \textbf{1) Intelligence Discovery}: this part focuses on the collection of software threat intelligence sources so that we can keep monitoring and tracking sources that are mostly used to publish such information, so that we can capture necessary software threat intelligence in time.
    \textbf{2) Knowledge Evolution Platform}: based on the identified intelligence sources, \tool iteratively conducts continuous monitoring, parsing, validating, and utilizing of software threat intelligence, and meanwhile, iterate the corresponding data model and understanding model of intelligence, threat, knowledge, and context, to evolve the capability of \tool along the time. 

Based on these two major parts, we expect not only \tool to be able to automate the capability of manipulating software threat intelligence, but also take software threat intelligence as an example to explore the solutions to automate the explicit knowledge collection, interpretation, cognition alignment, validation, and utilization for other domains.

\section{Intelligence Discovery}

First, \tool is initiated by existing software threat intelligence. Specifically, we rely on existing well-known and high quality sources as the initial trustworthy sources to collect software threat intelligence. For instance, information from sources like NVD, OSV, Snyk, GitHub Advisory, and VulnDB are usually with human validation, which are also widely used as the source data of software auditing tasks (i.e., software composition analysis) and with well-structured information that are easy to interpret. To this end, starting from them, \tool collects all references provided on these platforms for each identified software threat reports, and based on these references, \tool recursively extend to include possible reference links to of them to identify as many possible original sources of software threat intelligence as possible. Moreover, based on these collected reference software threat intelligence (i.e., webpages), \tool also refer to search engines for better coverage of intelligence sources by searching most commonly used terms in both all intelligence and single ones. After the extensive search, \tool then conduct relevance analysis to exclude those that are not related to the specific tasks (i.e., only focusing on vulnerabilities or malicious packages). Finally, \tool further identifies the software threat intelligence sources that are worth tracking as the initial sources to monitor.

\section{Knowledge Evolution via Multi-Agent Systems}

Next, we introduce the multi-agent system that self-evolves the interpretation, completion, validation, and utilization of software threat intelligence.

\textbf{1) Intelligence Interpretation.}
Starting from software threat intelligence \tool collected, \tool incorporates a multi-step solution to identify the key information that worth extracting and preserving. Specifically, it starts with identifying all entities mentioned in the free texts, and identifies the possible relationship among them. Note that although there could be many less relevant entities extracted and implicit relationships not covered, we take these results as the initial sets of key information. Next, for similar intelligences (i.e., same vulnerabilities or malicious packages), we align the entities and relationships among these results, and identify what are the key objects to focus on, and merge them into a complete knowledge graph for specific cases. As for the possible conflicts, \tool incorporates a guided reasoning based on all knowledge it has to determine the correct results. 
Meanwhile, after going through these identified key information of software threat intelligence, \tool also gradually formalizes an intelligence model as the meta model of all possible key information recorded in existing software threat intelligence.

\textbf{2) Intelligence Completion.}
Apart from the alignment of software threat intelligence, it is also possible that some key information identified in the intelligence model is not available in any intelligence for specific cases. To this end, \tool borrows the rationale of the missed information in other similar software threats. By analyzing the rationale in existing cases, and how this information is extracted, \tool follows the logic to inspect similar localisations (i.e., source code repositories) to check whether the missed information can be made up properly.
Moreover, this process also clarifies the threat model of specific cases, such as how the threats are formed, and how are they threaten corresponding user projects. Based on this, the common patterns of threat models can be concluded, to facilitate downstream validation and detection of new emerging software threat intelligence.

\textbf{3) Intelligence Validation.}
After completing the necessary vulnerability information, \tool further validates the existence and real threats of reported software security threats. Specifically, based on the threat model, \tool conducts the root cause analysis to identify the conditions and rationale of how the corresponding software threats threaten user projects. By generating the minimum context and constructing a runtime environment, \tool further migrates existing Proof-of-Concept (POC) or generate the POCs directly to trigger the threats and determine whether they have a real impact on user projects.
Based on this, \tool can also gradually formulate the knowledge of software threats, such as mappings of root causes and validation strategies, the ways to generate test cases and corresponding environment setups, etc. By concluding them, \tool gradually enriches the knowledge of threats.

\textbf{4) Software Risk Detection.}
Based on the threat knowledge, \tool can also stem the critical criteria to judge the existence of software threats, which can further enhance existing detection approaches. Specifically, software security threats usually involve sensitive implementations, while not all sensitive operations lead to real risks to user projects. To this end, \tool could further incorporate logical reasoning of LLMs to judge whether identified sensitive operations can lead to real impact. Based on them, \tool can derive not only a better refined explanation of software threats, from their rationale to presence, but also more precise detection rules that are with higher precision while fewer false positives. 
Moreover, these generated rules and explanations can also propose new key information that is vital for software threats, and they would be evolved back to the interpretation of software threat intelligence to better capture relevant information.

With these key steps and models, \tool can gradually evolve to better collect, interpret, complete, validate, and utilize software threat intelligence.

\textbf{5) Preliminary Results.}
\tool has demonstrated significant early-stage outcomes, validating the effectiveness and scalability of the proposed methodology. 
So far, \tool has integrated with over 30 authoritative and 100 open intelligence sources, which provide extensive threat coverage. Moreover, \tool has successfully manages over 260,000 vulnerabilities and has conducted in-depth analyses on more than 30,000 malicious code implementations, which is so far the largest open dataset of malicious packages~\cite{guo2024packageintelleveraginglargelanguage}. 
Furthermore, \tool has also integrated and compared 18+ different major sources of vulnerabilities, resulting in a unified framework of vulnerability key information presentation. These early vulnerability metadata modelling and source distribution analytics confirm the analytical strength and adaptability of \tool.


\tool represents a critical evolution in software intelligence management. By dynamically adapting through continuous intelligence refinement and validation, the platform provides significant improvements over traditional static intelligence approaches. Future work includes but is not limited to expanding agent capabilities, such as conflict detection in intelligence interpretation, the analogy of rationale among similar threats for completion, automated compilation and build of software for validation, and more efficient automated iterative refinement of rules for detection, etc.

\section{Conclusion}

In this paper, we propose the roadmap of \tool, which establishes a framework towards the automated self-evolving platform for the collection, interpretation, completion, validation, and utilization of software threat intelligence. Through sophisticated integration of intelligence gathering, analytical reasoning, and threat modelling, \tool delivers comprehensive, scalable, and dynamic security solutions capable of addressing complex, evolving software threats.

\section{Acknowledgment}
This research is supported by the Ministry of Education, Singapore, under its Academic Research Fund Tier 1 (RG96/23). It is also supported by the National Research Foundation, Singapore, and DSO National Laboratories under the AI Singapore
Programme (AISG Award No: AISG2-GC-2023-008); by the National Research Foundation Singapore and the Cyber Security Agency under the National Cybersecurity R\&D Programme (NCRP25-P04-
TAICeN); and by the Prime Minister’s Office, Singapore under the Campus for Research Excellence and Technological Enterprise (CREATE) programme. Any opinions, findings and conclusions, or
recommendations expressed in these materials are those of the author(s) and do not reflect the views of National Research Foundation, Singapore, Cyber Security Agency of Singapore, Singapore.

%
%
%
\bibliographystyle{splncs04}
\bibliography{ref}

\end{document}